%% file: main_arxiv.tex
\documentclass[conference]{IEEEtran}
\IEEEoverridecommandlockouts
\usepackage[letterpaper,top=0.74in,bottom=1in,left=0.625in,right=0.625in]{geometry}
\usepackage{amsmath,amssymb,amsfonts}
\usepackage[caption=false]{subfig}
\usepackage{mathtools} 
\usepackage{adjustbox} 
\usepackage{graphicx}
\usepackage{textcomp}
\usepackage{xcolor}
\usepackage{tabularx}
\usepackage{booktabs}
\usepackage{float}
\usepackage{algpseudocode}
\usepackage[english]{babel}
\usepackage{graphicx,lipsum}
\usepackage{paralist}
\usepackage{array}
\usepackage{amsmath,stackrel}
\usepackage{multicol}
\usepackage{verbatim}
\usepackage{enumerate}              
\usepackage{dsfont}
\usepackage{psfrag}
\usepackage{algorithm}
\usepackage[nolist]{acronym}
\usepackage[utf8]{inputenc}
\usepackage{tikz}
\usepackage{collcell}
\usepackage{pgfplots}
\pgfplotsset{compat=1.18}
\usepackage{hhline}
\usepackage{colortbl}
\usepackage{multido}
\usepackage{multirow}
\usepackage{etoolbox}
\usepackage{soul}
\usepackage{eucal,balance}

\input{acronyms.tex}

\input{macros.tex}

\def\undertilde#1{\mathord{\vtop{\ialign{##\crcr
$\hfil\displaystyle{#1}\hfil$\crcr\noalign{\kern1.5pt\nointerlineskip}
$\hfil\tilde{}\hfil$\crcr\noalign{\kern1.5pt}}}}}

\def\BibTeX{{\rm B\kern-.05em{\sc i\kern-.025em b}\kern-.08em
    T\kern-.1667em\lower.7ex\hbox{E}\kern-.125emX}}
\begin{document}

\title{Joint Target Acquisition and Refined Position Estimation in OFDM-based ISAC Networks}

\author{Lorenzo~Pucci and Andrea~Giorgetti
\thanks{Authors are with the Wireless Communications Laboratory (WiLab), CNIT, and the Department of Electrical, Electronic, and Information Engineering “Guglielmo Marconi" (DEI), University of Bologna, 40136, Italy.
Email: \{lorenzo.pucci3, andrea.giorgetti\}@unibo.it}
\thanks{This work was supported in part by the CNIT National Laboratory WiLab and the WiLab-Huawei Joint Innovation Center and in part by the European Union under the Italian National Recovery and Resilience Plan (NRRP) of NextGenerationEU, partnership on ``Telecommunications of the Future'' (PE00000001 - program ``RESTART'').}
}
\medskip

\maketitle

\begin{abstract}
This paper addresses joint target acquisition and position estimation in an OFDM-based integrated sensing and communication (ISAC) network with base station (BS) cooperation via a fusion center. A two-stage framework is proposed: in the first stage, each BS computes range-angle maps to detect targets and estimate coarse positions, exploiting spatial diversity. In the second stage, refined localization is performed using a cooperative maximum likelihood (ML) estimator over predefined regions of interest (RoIs) within a shared global reference frame. Numerical results demonstrate that the proposed approach not only improves detection performance through BS cooperation but also achieves centimeter-level localization accuracy, highlighting the effectiveness of the refined estimation technique.
\end{abstract}


\section{Introduction}
\Ac{ISAC} systems aim to jointly support data transmission and environmental sensing. While cascaded two-stage \ac{ISAC} frameworks have gained attention, few works have addressed both target detection and refined localization, particularly in cooperative settings. In \cite{Shietal22}, a non-cooperative, device-free scheme is proposed combining \ac{OFDM}-based range estimation and localization using measurements from multiple \acp{BS}. In \cite{Jafrietal25}, the authors incorporate detection into a two-stage design, estimating target presence and refining radar parameters in non-cooperative setups. Among cooperative designs, \cite{Zhangetal24} proposes a two-stage localization framework that first applies a 2D-\ac{FFT} for delay-Doppler extraction, followed by position estimation after filtering ill-conditioned measurements—yet they omit explicit detection. Similarly, in \cite{Liuetal24}, the authors adopt a multi-BS approach with a measurement association algorithm but assume the target is already detected. Despite such contributions, to the authors' knowledge, a fully cascaded framework integrating detection and position refinement in a cooperative setting is missing. 

To address this gap, we propose a complete two-stage framework for joint detection and refined localization in \ac{OFDM}-based \ac{ISAC} networks. In Stage~1, each \ac{BS} computes local range-angle maps via beamforming and periodogram to detect targets and provide coarse localization, with cooperation enabling spatial diversity gain. Stage~2 refines localization via a cooperative \ac{ML} estimator over a shared global reference system, limited to a predefined \ac{RoI} to manage complexity. Simulation results confirm the effectiveness of the proposed design, showing that \ac{BS} cooperation significantly improves both detection and localization accuracy (down to centimeter-level) with minimal additional complexity.

In this paper, bold capital/lowercase letters denote matrices/vectors; $(\cdot)^\transp$, $(\cdot)^\herm$, and $\|\cdot\|$ denote transpose, Hermitian, and Euclidean norm; $\mathbf{I}_n$ is the $n \times n$ identity matrix, $\mathrm{diag}(\cdot)$ is a diagonal matrix, and $\mathbb{E}\{\cdot\}$ is the expectation. The notations $\mathbf{x} \sim \mathcal{CN}(\mathbf{0},\boldsymbol{\Sigma})$ and $\mathbf{x} \sim \mathcal{N}(\boldsymbol{\mu},\boldsymbol{\Sigma})$ refer to circularly symmetric complex (zero mean) and real Gaussian vectors, respectively, with covariance $\boldsymbol{\Sigma}$. The operator $\angle\cdot$ denotes the phase, $|\cdot|$ is the absolute value or cardinality (depending on context), and $\otimes$ is the Kronecker product. Lastly, $\ceil{\cdot}$ is the ceiling function.


\section{System Model}\label{sec:system_model}
\begin{figure}[t]
    \centering
    \includegraphics[width=0.7\columnwidth]{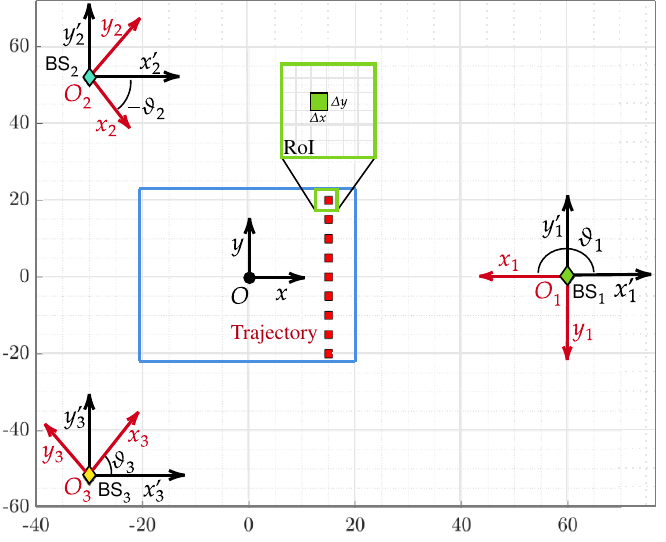}
    \caption{\ac{ISAC} network with multiple \acp{BS} performing sensing and communication. The red square indicates the target trajectory, the green square marks the refined estimation \ac{RoI}, and the blue square defines the monitored area.}
    \label{fig:scenario}
\end{figure}

We consider an \ac{OFDM}-based \ac{ISAC} network comprising multiple monostatic \acp{BS} coordinated via a \ac{FC} (see Fig.~\ref{fig:scenario}). The \ac{FC}  performs sensor fusion and manages interference through frequency and time division. Each \ac{BS} features two half-wavelength spaced \acp{ULA}, one with $N_\mathrm{T}$ antennas for joint communication and sensing transmission, and one with $N_\mathrm{R}$ receive antennas for sensing.

Sensing follows a two-stage process: (i) a target acquisition phase for detection and coarse localization, and (ii) a refinement phase for precise localization. Both stages run continuously,\footnote{Alternatively, the refinement stage can run continuously for tracking, while the acquisition stage is triggered periodically to detect new targets.} but are analyzed separately in this work.

In Stage~1, each \ac{BS} computes local range-angle maps using the same downlink data-carrying signal employed for communication. By precoding the transmitted symbols to form multibeam patterns, energy is effectively shared between sensing and communication directions. Range-angle maps are then obtained via a double periodogram across multiple beam directions \cite{PucPaoGio:J22}. In Stage~2, refined localization is performed using a cooperative grid-based \ac{ML} estimator, similar to \cite{PucBacGio:L25}, operating in a global Cartesian coordinate system shared by all \acp{BS}, enabling tighter cooperation. In this stage, digital beamforming is used to illuminate in a time-division manner specific \acp{RoI} where targets were previously detected in Stage~1. Details are provided in Section~\ref{sec:two_stage}.

\subsection{Transmit Signal for OFDM-based Sensing}
Each \ac{BS} transmits an \ac{OFDM} frame of duration $T_\mathrm{tot} = M T_\mathrm{s}$ and maximum bandwidth $B_\mathrm{max} = K \Delta f$ with carrier frequency $f_\mathrm{c} \gg B_\mathrm{max}$. Here, $M$ is the number of \ac{OFDM} symbols per frame, $\Delta f$ is the subcarrier spacing, and $K$ is the total number of active subcarriers. The symbol duration is $T_\mathrm{s} = 1/\Delta f + T_\mathrm{cp}$, where $T_\mathrm{cp}$ denotes the duration of the cyclic prefix, which is used to mitigate \ac{ISI}. 

Sensing in both stages is enabled by reserving portions of the frame’s bandwidth, time, and/or power. Specifically, $K_\mathrm{s} \leq K$ subcarriers and $M_\mathrm{s} \leq M$ \ac{OFDM} symbols are used for sensing in a given direction. To complete a scan of the environment in the first stage or cover multiple \acp{RoI} in the second one, a total of $N_\mathrm{dir} M_\mathrm{s}$ symbols are allocated, where $N_\mathrm{dir}$ is the number of sensing directions. The fractions of power and bandwidth used for sensing are denoted by $\rho_p$ and $\rho_\mathrm{f} = K_\mathrm{s}/K$, respectively. The transmitted signal for each direction is represented in the frequency domain by $\mathbf{X}_\mathrm{s} \in \mathbb{C}^{K_\mathrm{s} \times M_\mathrm{s}}$, whose entries $x_{k,m}$ are complex modulation symbols with $\mathbb{E}\{|x_{k,m}|^2\} = 1$ \cite{PucPaoGio:J22}.

Each transmitted symbol $x_{k,m}$ is precoded by a beamforming vector $\mathbf{w}_\mathrm{T} \in \mathbb{C}^{N_\mathrm{T} \times 1}$, thus obtaining the vector $\mathbf{x}[k,m] = \mathbf{w}_\mathrm{T} x_{k,m}$. The choice of $\mathbf{w}_\mathrm{T}$ depends on the sensing stage (coarse or refined), as detailed in the following subsections.

\subsection{Input-Output Relationship of the First Sensing Stage}
\label{sec:1st-stage}
In Stage~1, the beamforming vector is designed to form a multibeam radiation pattern, enabling simultaneous downlink communication with one or more \acp{UE} and angular scanning for sensing \cite{PucPaoGio:J22}. The angular range $[-\theta_0, \theta_0]$ is scanned in steps of $\Delta \theta = 2\theta_0/(N_\mathrm{dir}-1)$, transmitting $M_\mathrm{s}$ \ac{OFDM} symbols per direction.

The transmit beamformer, which follows a beam steering strategy, is given by
$\mathbf{w}_\mathrm{T} = \sqrt{\frac{\rho_\mathrm{p}P_\mathrm{avg}}{N_\mathrm{T}}
}\mathbf{a}\big(\theta_{\mathrm{s}}\big) + \sqrt{\frac{(1-\rho_\mathrm{p})P_\mathrm{avg}}{N_\mathrm{T}}
}\mathbf{a}\big(\theta_{\mathrm{c}}\big)$, 
where $\rho_\mathrm{p} \in [0,1]$ is the power fraction for sensing, and $P_\mathrm{avg} = P_\mathrm{T}/K$ is the per-subcarrier power over the total power $P_\mathrm{T}$.
Furthermore, $\mathbf{a}(\theta_{\mathrm{s}}) \in \mathbb{C}^{N_\mathrm{T} \times 1}$ and $\mathbf{a}(\theta_{\mathrm{c}}) \in \mathbb{C}^{N_\mathrm{T} \times 1}$ are the array response vectors associated to sensing and communication \ac{DoD}, $\theta_{\mathrm{s}}$ and $\theta_{\mathrm{c}}$, respectively \cite{PucPaoGio:J22}. Taking as a reference the middle point of the array, the spatial steering vector $\mathbf{a}(\theta)$ for a generic direction $\theta$, is given by 
%
$\mathbf{a}(\theta)=\left [e^{-\imath \pi\frac{(N_\mathrm{T}-1)}{2} \sin(\theta)},\dots, e^{\imath \pi \frac{(N_\mathrm{T}-1)}{2}  \sin(\theta)} \right]^\mathsf{T}.$

Under far-field line-of-sight propagation conditions with $L$ scatterers, the $\Nr \times \Nt$ two-way frequency-domain channel matrix at subcarrier $k$ and symbol $m$ is modeled as 
\begin{equation}
\label{eq:channel-matrix}
    \mathbf{H}[k,m] = \sum_{l = 1}^{L} \alpha_l e^{\imath \phi_l} e^{\imath2\pi m T_\mathrm{s} f_{\mathrm{D},l}}e^{-\imath 2\pi k \Delta f \tau_l} \mathbf{b}(\theta_l)\mathbf{a}^\mathsf{H}(\theta_l)
\end{equation}
where $\tau_l$, $f_{\mathrm{D},l}$, and $\theta_l$ denote the round-trip delay, Doppler shift, and \ac{DoA}/\ac{DoD} of the $l$th target, respectively. The $l$th channel gain is $\alpha_l = \sqrt{\frac{G_\mathrm{T}G_\mathrm{R} c^2 \sigma_l}{(4\pi)^3 f_\mathrm{c}^2 r_l^4}}$ and the phase shift is $\phi_l = -2\pi f_\mathrm{c}\tau_l$. Additionally, $G_\mathrm{T}$ and $G_\mathrm{R}$ are the single-element transmit and receive antenna gains, $c$ is the speed of light, while $\sigma_l$ and $r_l =c\tau_l/2$ are the \ac{RCS} and the distance of target $l$ from the \ac{BS}, respectively. The \ac{RCS} follows the Swerling~I model, i.e., it has an exponential distribution with expected value $\bar{\sigma_l} =\mathbb{E}\{\sigma_l\}$ \cite{swerling:54}. Notably, since \acp{BS} are spatially distributed, each observes a different scattering profile, enabling spatial diversity and enhancing detection performance \cite{Fishler_spatial_diversity}. The vector $\mathbf{b}(\theta_l) \in \mathbb{C}^{N_\mathrm{R} \times 1}$ is the array response at the sensing receiver.

At the receiver, the $N_\mathrm{R} \times 1$ vector of demodulated symbols for subcarrier $k$ and \ac{OFDM} symbol $m$ is given by\footnote{Without loss of generality, we assume that \ac{ISI}, \ac{ICI}, and \ac{SI} are effectively suppressed, making their contributions negligible compared to Gaussian noise \cite{FullDuplex}.} 
\begin{equation}
\mathbf{y}[k,m] = \mathbf{H}[k,m] \mathbf{x}[k,m] + \mathbf{n}[k,m] 
    \label{eq:rx_vec_1st}
\end{equation}
where $\mathbf{n}[k,m] \sim \mathcal{CN}(\mathbf{0},\sigma_\mathrm{N}^2 \mathbf{I}_{N_\mathrm{R}})$ is the 
noise vector with $\sigma_\mathrm{N}^2 = N_0 \Delta f$, where $N_0$ is the noise \ac{PSD}.

\subsection{Input-Output Relationship of the Second Sensing Stage} \label{sec:2ndStage}
In Stage~2, the beamforming strategy follows the approach in \cite{Friedlander:12}. For each detected target $p$, a \ac{RoI} centered at the estimated position and aligned with the estimated target direction $\hat{\theta}_p$ is identified. The transmit beamforming vector $\mathbf{w}_\mathrm{T}(\hat{\theta}_p)$ satisfies $\|\mathbf{w}_\mathrm{T}(\hat{\theta}_p)\|^2 = P_\mathrm{avg}$ and is designed to maintain approximately constant gain across the sector containing the \ac{RoI} while suppressing sidelobes elsewhere. 
This ensures minimal interference from other non-overlapping targets in the angular domain, i.e., $\gamma_{l,p} \approx 0$ for $\theta_l \neq \theta_p$, where $\gamma_{l,p} = \mathbf{a}^\mathsf{H}(\theta_l)\mathbf{w}_\mathrm{T}(\hat{\theta}_p)/\sqrt{P_\mathrm{avg}}$ is the beamforming gain for target $l$ when pointing toward $p$. 

Since each \ac{RoI} has a fixed, pre-determined area based on the scenario, the beamwidth depends on the \ac{RoI} location relative to the \ac{BS}. Specifically, the $3\,$dB beamwidth of $\mathbf{w}_\mathrm{T}(\hat{\theta}_p)$ is defined as $\Delta\Theta = \theta_\mathrm{max} - \theta_\mathrm{min}$, with $\theta_\mathrm{max}$ and $\theta_\mathrm{min}$ denoting the bounding directions of the \ac{RoI}. The minimum beamwidth is limited by array resolution, e.g., $\Delta\Theta_\mathrm{min} \approx 2/N_\mathrm{T}$ for a ULA with half-wavelength spacing.


Under these assumptions, the received signal in \eqref{eq:rx_vec_1st} can be reformulated more conveniently. Using the channel model in \eqref{eq:channel-matrix}, and defining the overall gain $h_{l} \triangleq \sqrt{P_\mathrm{avg}}\, \gamma_{l,p} \, \alpha_l e^{\imath \phi_l}$ when pointing towards the sensing direction $\hat{\theta}_p$, we stack the received samples across time, frequency, and antennas to obtain the $K_\mathrm{s} M_\mathrm{s} N_\mathrm{R} \times 1$ vector of received symbols
\begin{equation}
    \tilde{\mathbf{y}} = \sum_{l=1}^{L} h_{l}\mathbf{G}(\tau_l,f_{\mathrm{D},l},\theta_{l})\tilde{\mathbf{x}} + \tilde{\mathbf{n}}
   \label{eq:rx_blocked}
\end{equation}
where $\tilde{\mathbf{x}} \in \mathbb{C}^{K_\mathrm{s} M_\mathrm{s} \times 1}$ contains all transmitted sensing symbols $x_{k,m}$, and $\tilde{\mathbf{n}} \sim \mathcal{CN}(\mathbf{0}, \sigma_\mathrm{N}^2 \mathbf{I}_{K_\mathrm{s} M_\mathrm{s} N_\mathrm{R}})$ is the Gaussian noise vector. Moreover, we define the $K_\mathrm{s} M_\mathrm{s} N_\mathrm{R} \times K_\mathrm{s} M_\mathrm{s}$ effective channel matrix for a given scatterer $l$ as
\begin{equation} 
\mathbf{G} (\tau_l,f_{\mathrm{D},l},\theta_{l}) \triangleq \mathbf{b}(\theta_{l}) \otimes  \mathbf{T} (\tau_l,f_{\mathrm{D},l})
\label{eq:eff_ch_mat_1}
\end{equation}
where the matrix $\mathbf{T}(\tau_l, f_{\mathrm{D},l}) \in \mathbb{C}^{K_\mathrm{s}M_\mathrm{s}\times K_\mathrm{s}M_\mathrm{s}}$ is defined as
{\small
\begin{multline}
    \label{eq:OFDM_steer_vecs}
    \mathbf{T}(\tau_l, f_{\mathrm{D},l}) \triangleq 
    \mathrm{diag}([1, \dots, e^{\imath 2\pi m T_\mathrm{s} f_{\mathrm{D},l}}, \dots, e^{\imath 2\pi (M_\mathrm{s}-1) T_\mathrm{s} f_{\mathrm{D},l}}]^\mathsf{T}\\
    \otimes [1, \dots, e^{-\imath 2\pi k\Delta f\tau_l}, \dots, e^{-\imath 2\pi (K_\mathrm{s}-1)\Delta f \tau_l}]^\mathsf{T}).
\end{multline}}

Without loss of generality, we consider targets spatially separated—either in angle, delay, or both domains—so that only one target lies within each \ac{RoI}. Stage~2 ends after scanning $N_\mathrm{dir} = \hat{L} \leq L$ \acp{RoI}, where $\hat{L}$ is the number of targets detected in the first stage.

\section{Two-stage Detection and Estimation}
\label{sec:two_stage}

\subsection{Coarse Parameter Estimation Via Range-Angle Maps}
The first sensing stage performs environmental scanning via beam sweeping at each \ac{BS}. For each sensing direction $\theta_\mathrm{s}$, the received signal in \eqref{eq:rx_vec_1st} is collected and processed using a receive beamforming vector $\mathbf{w}_\mathrm{R} = \mathbf{b}(\theta_\mathrm{s})/\sqrt{N_\mathrm{R}}$. 
The resulting received symbols $y_{k,m} = \mathbf{w}_\mathrm{R}^\mathsf{H}\mathbf{y}[k,m]$ are then processed to eliminate the dependence on the transmitted ones, through element-wise division (also known as reciprocal filtering), yielding the matrix $\widetilde{\mathbf{Y}}$ with elements
$\widetilde{y}_{k,m}=\frac{y_{k,m}}{x_{k,m}}=\mathbf{w}_\mathrm{R}^\mathsf{H}\mathbf{H}[k,m]\mathbf{w}_\mathrm{T}+\frac{\mathbf{w}_\mathrm{R}^\mathsf{H}\mathbf{n}[k,m]}{x_{k,m}}$.\\
A double periodogram is then applied across time and frequency to generate a range-Doppler map $\mathcal{P}(q, p)$ \cite{PucPaoGio:J22, Braun}
\begin{equation}\label{eq:period}
    \mathcal{P} (q, p) =  \frac{1}{K_\mathrm{s}\mathrm{M_\mathrm{s}}}\left|\sum_{k=0}^{K_\mathrm{p}-1} \biggl( \sum_{m=0}^{M_\mathrm{p}-1} \widetilde{y}_{k,m} e^{-j2\pi \frac{mp}{M_\mathrm{p}}}\biggr)e^{j2\pi \frac{kq}{K_\mathrm{p}}}\right|^2
\end{equation}
where $q=0,\dots,K_\mathrm{p}-1$ and $p=0, \dots, M_\mathrm{p}-1$, with $M_\mathrm{p}\geq M_\mathrm{s}$ and $K_\mathrm{p}\geq K_\mathrm{s}$ accounting for the zero-padding factor applied to the Doppler and delay estimation, respectively.

To construct the range-angle map, we consider that each sensing direction $j=0,\dots,N_\mathrm{dir}-1$ is associated with at most one dominant scatterer; an assumption that holds at \ac{mmWave} frequencies due to the achievable narrow beamwidths. Under this condition, the $j$th entry of the range-angle map $\mathcal{R}$ is obtained by selecting the highest-value column $p^j_\mathrm{max}$ in $\mathcal{P}$.
At each \ac{BS}, a hypothesis test is performed on the computed range-angle map $\mathcal{R}$ for all possible index pairs $(\bar{q}, j)$ to detect potential targets against noise. 
Each $(\bar{q}, j)$ index pair is uniquely associated to a $(r, \theta)$ pair as: $r =  \frac{\bar{q}\,c}{2\Delta f K_\mathrm{p}}$, and $\theta = -\theta_0 + j \Delta \theta$. Considering a maximum round-trip propagation delay $\tau \leq T_\mathrm{cp}$ to avoid \ac{ISI}, the index $\bar{q}$ vary over the range $\bar{q} = 0, \dots, \ceil{T_\mathrm{cp}\,\Delta f\,K_\mathrm{p}}$.

The detection problem is formulated as a binary hypothesis test, thus comparing range-angle maps against a threshold $\eta$
\begin{equation} \label{eq:likelihood-test}
   \mathcal{R}(\bar{q},j) = \mathcal{P}(\bar{q}, p^j_\mathrm{max})\overset{\mathcal{H}_0} {\underset{\mathcal{H}_1}{\lessgtr}} \eta
\end{equation}
where $\mathcal{H}_0$ and $\mathcal{H}_1$ denote the null (no target) and alternative (target present) hypotheses, respectively. To control the \ac{FAR} per map, the threshold $\eta$ is set based on the size of the discrete search space $\Omega$ (i.e., the set of all considered $(\bar{q}, j)$ pairs) as \cite{Braun}: 
$\eta = -\sigma^2_\mathrm{N} \ln(\textrm{FAR}/|\Omega|).$

Detections satisfying $\mathcal{R}(\bar{q}, j) > \eta$ are sent to the \ac{FC}, where clustering (e.g., DBSCAN \cite{EstKriXia:96}) is used to identify groups of detections for localization.  In specific scenarios, such as a single point-like target, localization can be simplified by directly identifying the global maximum (if $\geq \eta$) of the range-angle map at each \ac{BS}: $(\hat{\bar{q}}, \hat{j}) = \arg\max_{(\bar{q},j)} \mathcal{R}(\bar{q},j)$, which translates into the polar coordinates $(\hat{r}, \hat{\theta}) = \bigl(\frac{\hat{\bar{q}}\,c}{2\Delta f K_\mathrm{p}}, -\theta_0 + \hat{j} \Delta \theta \bigr)$.\footnote{Range-angle maps are expressed in local polar coordinates, which vary across \acp{BS}. Each $(r, \theta)$ pair refers to a point in the local reference frame, with the relationship between local and global coordinates discussed in Section~\ref{sec:coop_ML}.}

\begin{figure}[t]
    \centering
    \subfloat[Coarse range-angle map]{\includegraphics[width=.48\columnwidth] {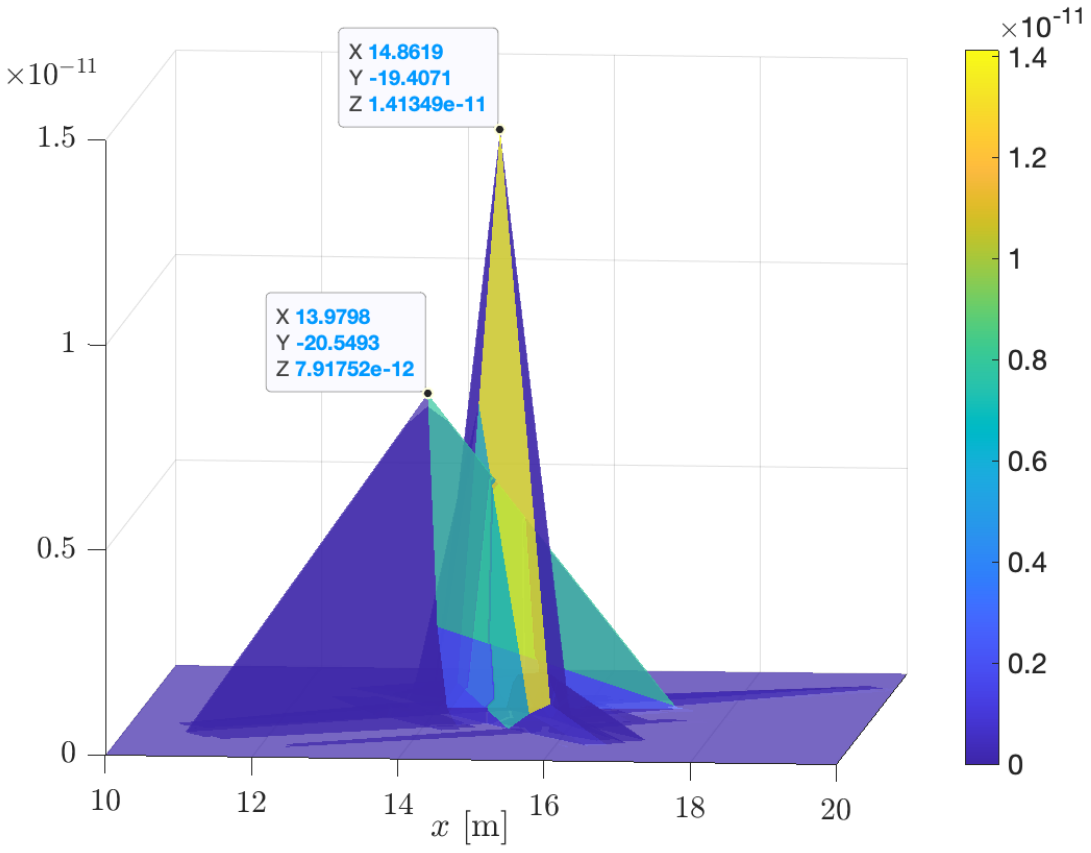}
    \label{fig:subfig1}} 
   \subfloat[Fine-grained likelihood map]{\includegraphics[width=.48\columnwidth]{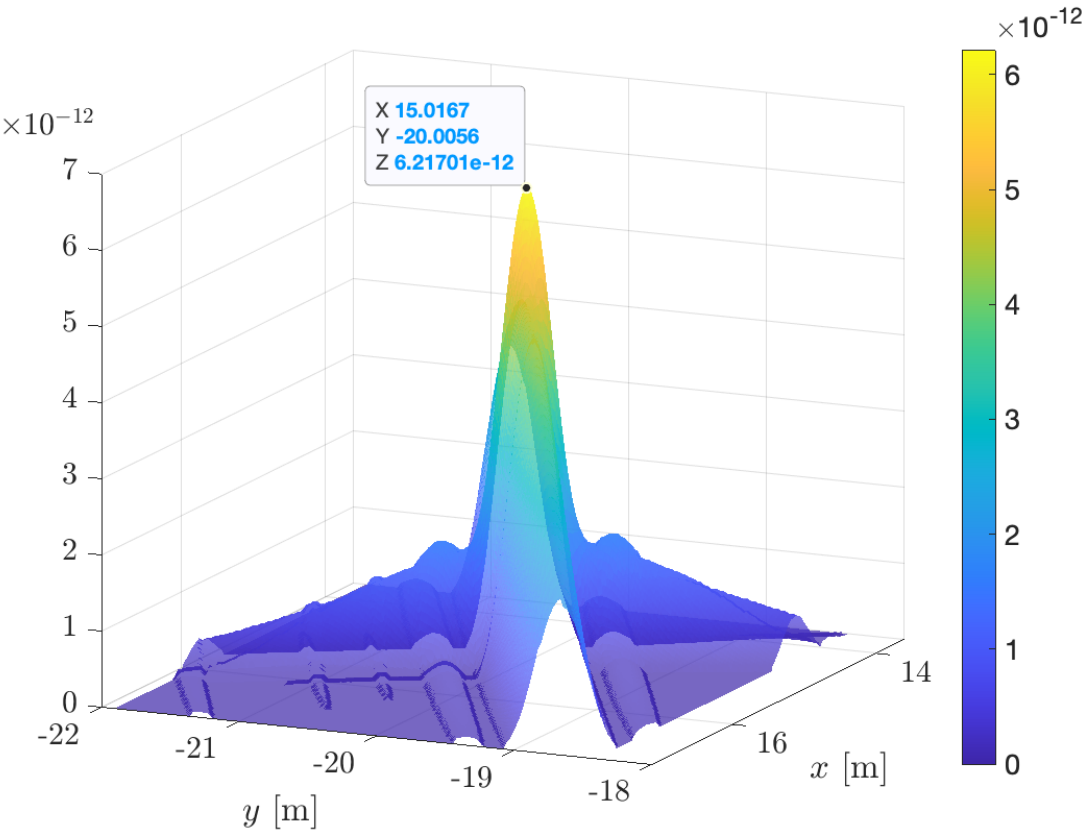}
   \label{fig:subfig2}}
    \caption{(a) Coarse and (b) refined cooperative maps for a target at $(15,-20)\,$m. Superimposed range-angle maps in (a) yield multiple peaks; the refined \ac{ML} map in (b) achieves accurate single-peak localization.}
    \label{fig:radar-map}
\end{figure}

Sensor fusion during this stage is performed using three techniques, described in Section~\ref{sec:num_res}.
An example of a fused range-angle map focused on a target area is shown in Fig.~\ref{fig:radar-map}(a). This map is obtained by superimposing the individual maps computed separately at each \ac{BS}. Multiple peaks emerge due to the non-perfectly aligned spatial sampling on independent \acp{BS}.

\subsection{Refined Cooperative Maximum Likelihood Estimation} \label{sec:coop_ML}

We consider the scenario in Fig.~\ref{fig:scenario}, with $\Nbs$ \acp{BS} located at global positions $\boldsymbol{\mathcal{O}}_i = [x^{(i)}_{\mathrm{bs}}, y^{(i)}_{\mathrm{bs}}]^\transp$, cooperating to localize $L$ point-like targets with positions $\mathbf{p}_l = [x_l, y_l]^\transp$ in a shared Cartesian reference system. 
The local coordinates of target $l$ relative to \ac{BS} $i$ are $\mathbf{p}_{l,i} =  [x_{l,i}, y_{l,i}]^\transp = [r_{l,i} \cos(\theta_{l,i}), r_{l,i} \sin(\theta_{l,i})]^\transp$, where $r_{l,i} = c\tau_{l,i}/2$. The relationship between the unknown local parameters $(\tau_{l,i}, \theta_{l,i})$ and the position $\mathbf{p}_l$ in global coordinates is \cite{PucBacGio:L25}
\begin{equation}
\label{eq:cartesian_to_polar}
\left\{\,\tau_{l,i}=(2/c)\sqrt{x^2_{l,i}+ y^2_{l,i}}; \quad \theta_{l,i}=
\arctan\left(y_{l,i}/x_{l,i}\right) \,\right\} 
\end{equation}
with $x_{l,i} = x'_{l,i}\cos(\vartheta_i) + y'_{l,i}\sin(\vartheta_i)$ and $y_{l,i} = -x'_{l,i}\sin(\vartheta_i) + y'_{l,i}\cos(\vartheta_i)$. Here, $x'_{l,i}=x_l-x^{(i)}_\mathrm{bs}$, $y'_{l,i}=y_l-y^{(i)}_\mathrm{bs}$, and $\vartheta_i \in (-\pi, \pi]$ is the rotation angle of the $i$th reference system. 

For each \ac{RoI}, the received vector $\tilde{\mathbf{y}}_i$ from \eqref{eq:rx_blocked} is collected at each \ac{BS}. Subsequently, a single vector of size $K_\mathrm{s} M_\mathrm{s} \Nr \Nbs \times 1$ is obtained as: $\underline{\mathbf{y}} = \bigl[\tilde{\mathbf{y}}_1^\transp, \tilde{\mathbf{y}}_2^\transp, \dots, \tilde{\mathbf{y}}^\transp_{\Nbs}\bigr]^\transp$.

The $l$th target parameter vector to be estimated relative to \ac{BS} $i$ is $\Thetab_{l,i} = [\beta_{l,i}, \varphi_{l,i}, f_{\mathrm{D},l,i}, \tau_{l,i},  \theta_{l,i}]^\transp$, where $\beta_{l,i} = |h_{l,i}|$ and $\varphi_{l,i}~=~\angle h_{l,i}$. The overall unknown parameter vector is
%
$\boldsymbol{\Xi}_l = \bigl[\Thetab_{l,1}^\transp, \Thetab_{l,2}^\transp \cdots, \Thetab^{\transp}_{l,\Nbs}\bigr]^\transp \in \Gamma$,
%
where $\Gamma = \mathbb{C}^{\Nbs} \times \mathbb{R}^{3 \Nbs}$ is the considered parameter space.

Assuming well-separated targets (cf. Section~\ref{sec:2ndStage}), we treat each \ac{RoI} as a single-target estimation problem and drop the subscript $l$ for clarity.  By ignoring those terms that are not relevant for the estimation and considering \eqref{eq:cartesian_to_polar}, the log-likelihood function of the received signal $\underline{\mathbf{y}}$ is given by  
\begin{equation}
    \label{eq:log-likelihood}
    l(\underline{\mathbf{y}};\boldsymbol{\Xi},\tilde{\mathbf{x}}) \approx -\frac{1}{\sigma^2_\mathrm{N}} \sum_{i=1}^{\Nbs} \bigl\| \tilde{\mathbf{y}}_i - h_i \mathbf{G}\Bigl(f_{\mathrm{D},i},\tau_i(\mathbf{p}),\theta_i(\mathbf{p})\Bigr) \tilde{\mathbf{x}}_i  \bigr\|^2
\end{equation}
where,  the parameters $\{\tau_i, \theta_i\}^{\Nbs}_{i=1} \subset \boldsymbol{\Xi}$ are expressed as a function of the target position $\mathbf{p}$ in the common reference system, i.e., $\{\tau_i(\mathbf{p})$, $\theta_i(\mathbf{p})\}^{\Nbs}_{i=1}$. 
Subsequently, an \ac{ML} estimation problem of the vector of unknowns $\boldsymbol{\Xi}$ can be formulated as: 
%
$\label{eq:ML_estim_multi}
\widehat{\boldsymbol{\Xi}}_\mathrm{ML} = \underset{\boldsymbol{\Xi} \in \Gamma}{\arg \max} \; l(\underline{\mathbf{y}};\boldsymbol{\Xi},\mathbf{x}).$

To reduce the problem to position-only estimation, we replace the unknown channel gains $h_i$ with their \ac{ML} estimates provided in \cite[eq. (21)]{PucBacGio:L25}.
Substituting $\widehat{h}_i$ into \eqref{eq:log-likelihood} and recalling the definition of $\mathbf{G}$ from \eqref{eq:eff_ch_mat_1} the cooperative \ac{ML} estimator is
\begin{equation}
    \widehat{\mathbf{p}} = \underset{\mathbf{p} \in \mathbb{R}^2}{\arg\max} \; \sum_{i=1}^{\Nbs} \frac{\left|\Bigl(\mathbf{W}\bigl(\theta_i(\mathbf{p})\bigr) \tilde{\mathbf{y}}_i\Bigr)^\mathsf{H} \mathbf{T}\Bigl(\tau_i(\mathbf{p}),\widehat{f}_{\mathrm{D},i}\Bigr)\tilde{\mathbf{x}}_i\right|^2}{\left\|\mathbf{T}\bigl(\tau_i(\mathbf{p}),\widehat{f}_{\mathrm{D},i}\bigr) \tilde{\mathbf{x}}_i\right\|^2} 
\label{eq:ML_pos_estim}
\end{equation}
where $\widehat{f}_{\mathrm{D},i}$ is from coarse estimation or set to a nominal value (e.g., $0\,$Hz) when Doppler resolution is poor (due to a low number of OFDM symbols), and $\mathbf{W}(\theta_i(\mathbf{p})) = \mathbf{w}(\theta_i(\mathbf{p})) \otimes \mathbf{I}_{K_\mathrm{s} M_\mathrm{s}}$, where $\mathbf{w}(\theta_i(\mathbf{p})) = \mathbf{b}^\mathsf{H}(\theta_i(\mathbf{p}))/\sqrt{N_\mathrm{R}}$ is a $1 \times N_\mathrm{R}$ unit-norm beamformer. To suppress sidelobes, Dolph–Chebyshev windowing may be applied \cite{Friedlander:12}.

Equation \eqref{eq:ML_pos_estim} is solved numerically by discretizing the 2D \ac{RoI} into a grid of $(x,y)$ points with steps $\Delta x$ and $\Delta y$. For each grid point $\mathbf{p}$, beamforming toward $\theta_i(\mathbf{p})$ is applied. 
Since nearby grid points yield nearly identical angles (up to three decimal digits in rad), beamforming can be reused across several points, reducing computation. Moreover, to further reduce complexity, $\mathbf{T}$ is modeled as a sparse diagonal matrix.

\input{table}


Notably, \eqref{eq:ML_pos_estim} forms a sum of $\Nbs$ likelihood maps over the $x$-$y$ plane, enabling two strategies: (i) each \ac{BS} computes and sends its map to the \ac{FC}, or (ii) raw received signals are sent to the \ac{FC}, which computes the global map. An example of fused map is shown in Fig.~\ref{fig:radar-map}(b).
\begin{figure*}[t]
    \centering
     \subfloat[Detection Probability]{\includegraphics[width=.3\linewidth]{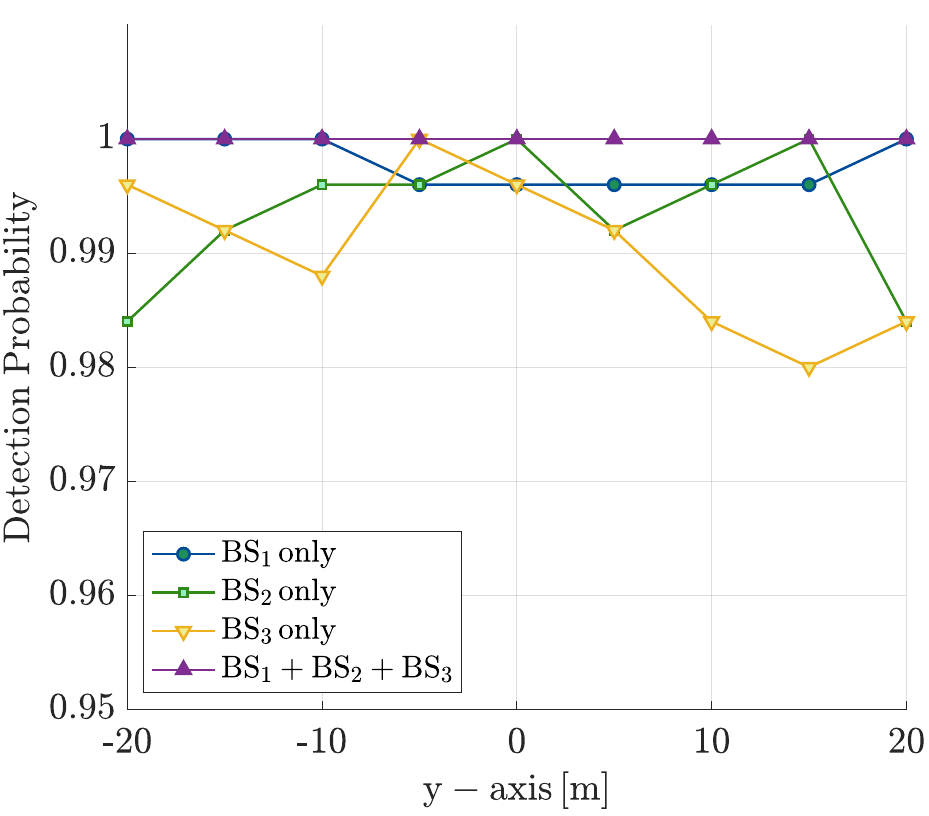}
     \label{fig:subfig1}
     }
     \qquad
    \subfloat[Single-\ac{BS} coarse position estimation]{\includegraphics[width=.3\linewidth]{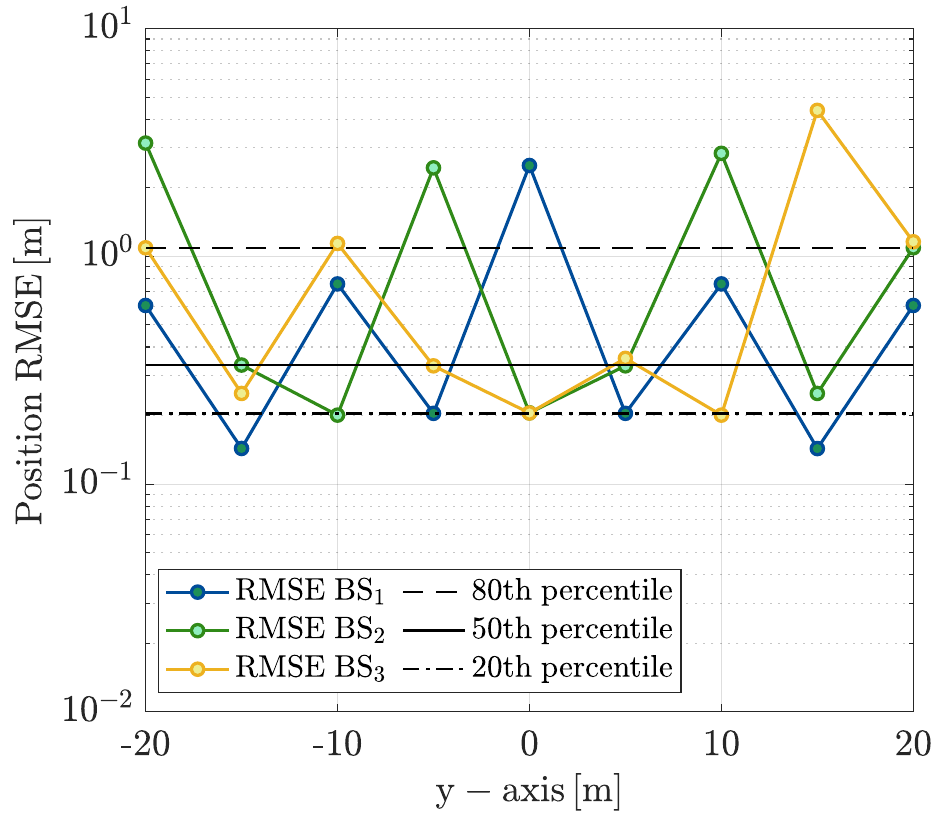}
    \label{fig:subfig2}
    }
    \qquad
    \subfloat[Cooperative position estimation]{\includegraphics[width=.3\linewidth]{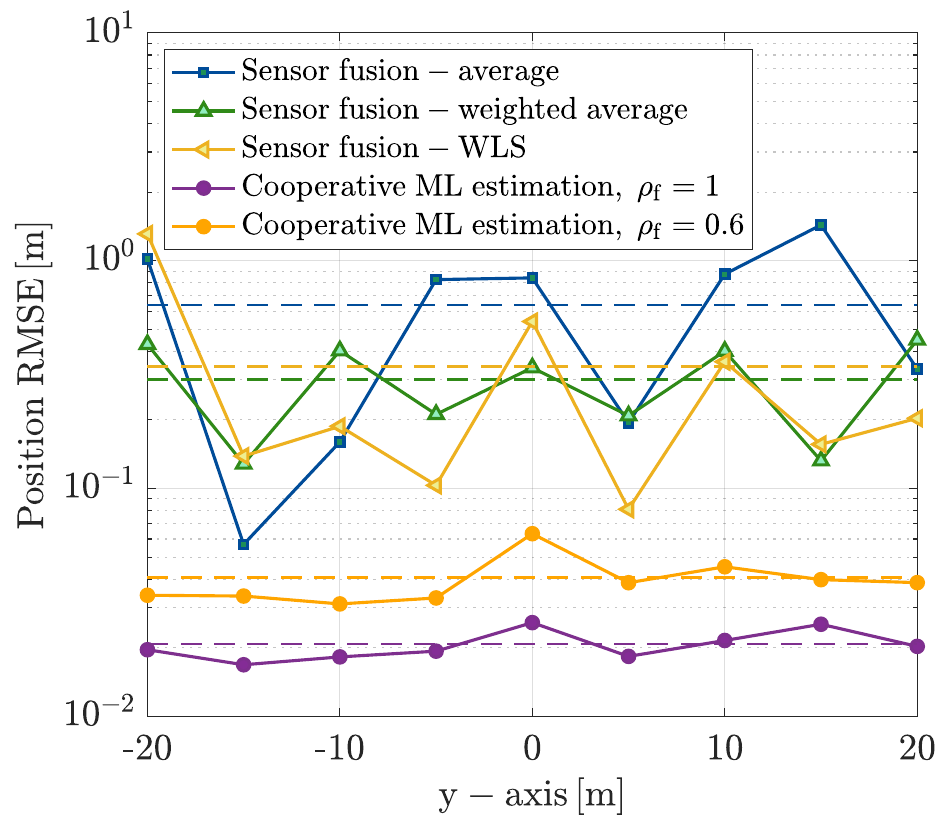}
    \label{fig:subfig3}
    }
    \qquad
    \caption{Detection probability and position RMSE: (a) Detection probability for single-\ac{BS} vs. cooperative setups. (b) Single-BS RMSE and error percentiles. (c) Cooperative RMSE for coarse/refined stages using different fusion techniques and bandwidth fractions. Dashed lines in (c) show mean RMSEs.}
    \label{fig:num_res}
\end{figure*}
\section{Numerical Results} \label{sec:num_res} 
The proposed two-stage detection and estimation framework is validated through simulations in the scenario shown in Fig.~\ref{fig:scenario}, with three \acp{BS} located at $\boldsymbol{\mathcal{O}}_1 = [60, 0]^\transp$, $\boldsymbol{\mathcal{O}}_2 = [-30, 52]^\transp$, and $\boldsymbol{\mathcal{O}}_3 = [-30, -52]^\transp$, oriented at angles $\boldsymbol{\vartheta} = [\pi, -\pi/3, \pi/3 ]^\transp$.

We evaluate performance for a single point-like target with mean \ac{RCS} $\bar{\sigma} = 1\,$m$^2$, moving along the trajectory in Fig.~\ref{fig:scenario} with fixed $x = 15\,$m. Simulations employ 5G NR-compliant signals with \ac{QPSK} modulation. Key parameters are listed in Table~\ref{tab:sim_param}. During Stage~1, range-Doppler maps are computed at each \ac{BS} for each sensing direction (with  $\theta_0=60^\circ$), using $K_\mathrm{p} = 4096$ and $M_\mathrm{p} = 256$, yielding range-angle maps of size $289 \times 50$. Around $10\,$ms are necessary to complete the angular scan. During Stage~2, likelihood maps are generated over $16\,\text{m}^2$ \acp{RoI} with resolution $\Delta x = \Delta y = 2\,$cm. The influence of the first-stage localization on the second stage is taken into account by modeling the \ac{RoI} center as $\hat{\mathbf{p}} \sim \mathcal{N}(\mathbf{p}, \tfrac{\sigma_\mathbf{p}^2}{2}\mathbf{I}_2)$, where $\sigma_\mathbf{p}$ is the coarse localization \ac{RMSE} and $\mathbf{p}$ the true position.

All results are averaged over $N_\mathrm{iter} = 250$ Monte Carlo trials per trajectory point, focusing on: (i) detection probability evaluated with and without \ac{BS} cooperation at fixed $\mathrm{FAR}=10^{-3}$. During cooperation, a detection is declared if at least one \ac{BS} observes the target. (ii) localization accuracy assessed via the \ac{RMSE} over successful detections, $N_\mathrm{det}$, both in single-\ac{BS} and cooperative settings. The \ac{RMSE} is computed as 
$\text{RMSE} = \sqrt{\tfrac{1}{N_\mathrm{det}}\sum_{j=1}^{N_\mathrm{det}}\bigl\| {\mathbf{p}}_j - {\widehat{\mathbf{p}}}_j \bigr\|^2}$, 
where $\mathbf{p}_j$ and $\hat{\mathbf{p}}_j$ denote true and estimated target positions, in local or global coordinates depending on the context.

\subsection{Detection Probability and Localization Accuracy}
Detection performance is evaluated during Stage~1, where each \ac{BS} independently detects the target based on whether the corresponding peak in the range-angle map $\mathcal{R}$ exceeds the threshold in \eqref{eq:likelihood-test}. 

Fig.~\ref{fig:num_res}(a) shows the detection probability across the trajectory, computed as the ratio $N_\mathrm{det}/N_\mathrm{iter}$ over Monte Carlo trials. In single-\ac{BS} scenarios, detection probability varies between $0.98$ and $1$. In contrast, with cooperation, it remains consistently at $1$, confirming that the target is always detected by at least one \ac{BS} in our scenario. This confirms the spatial diversity gain enabled by cooperative sensing.

Fig.\ref{fig:num_res}(b) and Fig.\ref{fig:num_res}(c) show the position RMSE along the target trajectory for the single-\ac{BS} and cooperative cases, respectively. In Fig.~\ref{fig:num_res}(c), dashed lines represent the average RMSE across the trajectory. For the cooperative setup, three sensor fusion methods are evaluated during the coarse localization stage, followed by refined localization using the estimator in \eqref{eq:ML_pos_estim}. The impact of bandwidth is also considered.

In the single-\ac{BS} case (Fig.~\ref{fig:num_res}(b)), localization is affected by quantization errors from the coarse range-angle resolution, resulting in unwanted oscillations in performance along the trajectory. To characterize error statistics, we report the $20$th, $50$th, and $80$th percentiles of the position error $e =\|\mathbf{p}-\hat{\mathbf{p}}\|$, aggregated across all \acp{BS}, points, and trials. The error remains below $1\,$m in 80\% of the cases, which is adequate for coarse tracking.

For the cooperative case in Fig.~\ref{fig:num_res}(c), three fusion methods are considered during coarse localization: i) \textit{Simple average} where estimates from each \ac{BS} are remapped to global coordinates and averaged, using only the \acp{BS} that detected the target. ii) \textit{Weighted average} where estimates are weighted by the peak intensity $\mathcal{R}^i_{\mathrm{max}}$ of the range-angle map at each \ac{BS} $i$, previously normalized by the global maximum across peak values from all the \acp{BS}. iii) \textit{WLS method} where a linear system is constructed from angle and range estimates and solved via \ac{WLS} as in \cite{PucBacGio:L25}. Precisely, the \ac{WLS} solution is obtained according to \cite[eq. (20)]{ding2025LsRssAoaBased}, with range replacing RSS measurements. The weighting matrix $\mathbf{W} =\mathrm{diag}(\mathcal{R}^1_{\mathrm{max}}, \dots, \mathcal{R}^1_{\mathrm{max}} \cdot \mathcal{R}^i_{\mathrm{max}}, \mathcal{R}^i_{\mathrm{max}}, ..., \mathcal{R}^1_{\mathrm{max}} \cdot \mathcal{R}^N_{\mathrm{max}}, \mathcal{R}^N_{\mathrm{max}}) \in \mathbb{R}^{(2N-1)\times (2N-1)}$ encodes reliability, with $\mathcal{R}^1_{\mathrm{max}}$ as the normalized weight of the reference \ac{BS} (here selected as the one with the highest map peak intensity), and $N$ the number of \acp{BS} detecting the target. The method requires $N \geq 2$ for a valid solution. Both the weighted average and \ac{WLS} yield similar mean \acp{RMSE}, $\sim30\,$cm, significantly outperforming the simple average, which settles to $\sim63\,$cm.

Refined localization is then performed assuming a coarse accuracy of $\sigma_\mathbf{p} = 70\,$cm, achieving mean \acp{RMSE} of  $4\,$cm and $2\,$cm for $\rho_\mathrm{f} = 0.6$ and $1$, respectively—up to $15\times$ improvement over coarse estimates. Notably, the finer spatial grid effectively mitigates quantization artifacts. 
\section{Conclusion}\label{sec:conclusion}
This work proposed a two-stage cooperative framework for joint detection and localization in \ac{OFDM}-based \ac{ISAC} networks. Stage~1 uses conventional periodogram-based range-angle maps for detection and coarse localization, while Stage~2 applies cooperative ML estimation for refined positioning. Numerical results have proven the effectiveness of the proposed two-step approach in achieving reliable detection and centimeter-level localization with up to $15\times$ accuracy improvement.

\bibliographystyle{IEEEtran}
\bibliography{IEEEabrv,bibliography}

\end{document}

%% file: acronyms.tex
\begin{acronym}
\acro{AEB}{angle error bound}
\acro{AWGN}{additive white Gaussian noise}
\acro{AoA}{angle of arrival}
\acro{AoD}{angle of departure}
\acro{AI}{artificial intelligence}
\acro{AIC}{Akaike information criterion}
\acro{BF}{beamformer}
\acro{BS}{base station}
\acro{CIR}{channel impulse response}
\acro{CP}{cyclic prefix}
\acro{CRB}{Cram\'er-Rao Bound}
\acro{CKF}{cubature Kalman filter}
\acro{CNN}{convolutional neural network}
\acro{CPHD}{cardinalized probability hypothesis density}
\acro{DBSCAN}{density-based spatial clustering of applications with noise}
\acro{DoA}{direction of arrival}
\acro{DoD}{direction of departure}
\acro{DFT}{discrete Fourier transform}
\acro{EF}{excision filtering}
\acro{EIRP}{effective isotropic radiated power}
\acro{ELP}{equivalent low-pass}
\acro{ESPRIT}{Estimation of Signal Parameters via Rotational Invariance Techniques}
\acro{FAR}{false alarm rate}
\acro{FFT}{fast Fourier transform}
\acro{FIM}{Fisher Information Matrix}
\acro{FD}{frequency division}
\acro{FC}{fusion center}
\acro{FR}{frequency range}
\acro{FDD}{frequency-division duplexing}
\acro{FoV}{field of view}
\acro{GM}{Gaussian mixture}
\acro{GDOP}{geometric dilution of precision}
\acro{GMPHD}{Gaussian mixture probability hypothesis density}
\acro{GMCPHD}{Gaussian mixture cardinalized probability hypothesis density}
\acro{GLRT}{generalized likelihood ratio test}
\acro{GOSPA}{generalized optimal sub-pattern assignment}
\acro{ICI}{inter-carrier interference}
\acro{IoT}{internet of things}
\acro{InF}{indoor factory}
\acro{IFFT}{inverse fast Fourier transform}
\acro{ISI}{inter-symbol interference}
\acro{ISAC}{integrated sensing and communication}
\acro{i.i.d.}{independent, identically distributed}
\acro{JSC}{joint sensing and communication}
\acro{k-NN}{k-nearest neighbors}
\acro{GLLR}{generalized log-likelihood ratio}
\acro{LLR}{log-likelihood ratio}
\acro{LOS}{line-of-sight}
\acro{LTE}{long term evolution}
\acro{MAP}{maximum a posteriori}
\acro{MB}{multi-Bernoulli}
\acro{MBM}{multi-Bernoulli mixture}
\acro{MCL}{maximum coupling loss}
\acro{MHT}{multiple hypothesis tracking}
\acro{MPL}{maximum path loss}
\acro{MSE}{Mean Square Error}
\acro{MIL}{maximum isotropic loss}
\acro{MC}{Monte Carlo}
\acro{mMIMO}{massive multiple-input multiple-output}
\acro{mmWave}{millimeter-wave}
\acro{MIMO}{multiple-input multiple-output}
\acro{MISO}{multiple-input single-output}
\acro{MUSIC}{MUltiple SIgnal Classification}
\acro{MDL}{minimum description length}
\acro{ML}{maximum likelihood}
\acro{NN}{neural network}
\acro{NR}{new radio}
\acro{NLoS}{non-line-of-sight}
\acro{OFDM}{orthogonal frequency division multiplexing}
\acro{OSPA}{optimal sub-pattern assignment}
\acro{p.d.f.}{probability density function}
\acro{PHD}{probability hypothesis density}
\acro{PSD}{power spectral density}
\acro{PCA}{principal component analysis}
\acro{PPP}{Poisson Point Process}
\acro{PEB}{position error bound}
\acro{PF}{particle filter}
\acro{PSK}{phase shift keying}
\acro{QAM}{quadrature amplitude modulation}
\acro{QPSK}{quadrature phase shift keying}
\acro{QuaDRiGa}{QUAsi Deterministic RadIo channel GenerAtor}
\acro{RCS}{radar cross-section}
\acro{REB}{range error bound}
\acro{ReLu}{rectified linear unit}
\acro{RF}{radio frequency}
\acro{RMSE}{root mean squared error}
\acro{RoI}{region of interest}
\acro{r.v.}{random variable}
\acro{RRH}{remote radio head}
\acro{RFS}{random finite set}
\acro{Rx}{receiver}
\acro{SSIR}{signal-to-self interference ratio}
\acro{SI}{self interference}
\acro{SIC}{successive interference cancellation}
\acro{SNR}{signal-to-noise ratio}
\acro{SU}{single-user}
\acro{SCM}{sample covariance matrix}
\acro{SPF}{soft particle filter}
\acro{TD}{time division}
\acro{TDD}{time-division duplexing}
\acro{ToA}{time of arrival}
\acro{TDoA}{time difference of arrival}
\acro{Tx}{transmitter}
\acro{THz}{terahertz}
\acro{U6G}{upper 6 GHz}
\acro{UMi}{urban micro}
\acro{UE}{user equipment}
\acro{ULA}{uniform linear array}
\acro{V2V}{vehicle-to-vehicle}
\acro{WGDOP}{weighted geometric dilution of precision}
\acro{WLS}{weighted least squares}
\end{acronym}

%% file: macros.tex
\newcommand{\Nbs}{N_\mathrm{bs}}

\newcommand{\transp}{\mathsf{T}}
\newcommand{\herm}{\mathsf{H}}

\newcommand{\Thetab}{\boldsymbol{\Theta}}

\newcommand{\Nt}{N_\mathrm{T}}
\newcommand{\Nr}{N_\mathrm{R}}

\DeclarePairedDelimiter\ceil{\lceil}{\rceil}

%% file: table.tex
\begin{table}[t]
\caption{Simulation Parameters (5G NR, FR2)}\label{tab:sim_param}
\centering
\renewcommand{\arraystretch}{1}
\begin{adjustbox}{width=0.88\columnwidth}
\begin{tabular}{l|c|c|c}
\toprule
\textbf{Parameter} & \textbf{Common} & \textbf{1st Stage} & \textbf{2nd Stage} \\
\midrule
Antennas ($N_\mathrm{T}, N_\mathrm{R}$) & $50$ &  &  \\
Carrier frequency $f_\mathrm{c}$ & $28\,$GHz &  &  \\
Subcarrier spacing $\Delta f$ & $120\,$kHz &  &  \\
Symbol duration $T_\mathrm{s}$ & $8.92\,\mu$s &  &  \\
OFDM symbols/frame $M$ & $1120$ &  &  \\
Active subcarriers $K$ & $3168$ &  &  \\
Power per subcarrier $P_\mathrm{avg}$ & $-5\,$dBm &  &  \\
Noise PSD $N_0$ & $4 \cdot 10^{-20}\,$W/Hz &  &  \\
Sensing directions $N_\mathrm{dir}$ &  & $50$ & $1$ \\
Symbols per direction $M_\mathrm{s}$ &  & $22$ & $1$ \\
Sensing subcarrier ratio $\rho_\mathrm{f}$ &  & $1$ & $1$, $0.6$ \\
Sensing power ratio $\rho_\mathrm{p}$ &  & $0.1$ &  $1$ \\
\bottomrule
\end{tabular}
\end{adjustbox}
\end{table}